\documentclass[epj]{svjour}
\usepackage{amssymb,epsfig}
\begin{document}

\title{
Searching for energetic cosmic axions in a laboratory experiment:\\
 testing the PVLAS anomaly
}
\titlerunning{
Searching for energetic cosmic axions in a laboratory experiment}
\author{Malcolm Fairbairn\inst{1} \and
Sergei Gninenko\inst{2} \and Nikolai Krasnikov\inst{2} \and Victor
Matveev\inst{2} \and
Timur Rashba\inst{3,4} \and Andre Rubbia\inst{5} \and Sergey
Troitsky\inst{2}}
\authorrunning{M.~Fairbairn et al.}
\institute{CERN, Geneva, Switzerland \and
Institute for Nuclear Research of RAS, 60th October
Anniversary Prospect 7a, Moscow 117312, Russia
\and
Max-Planck-Institut
f\"ur Physik (Werner-Heisenberg-Institut), F\"ohringer Ring 6, D-80805
M\"unchen, Germany
\and
Pushkov Institute
(IZMIRAN), Troitsk 142190, Russia
\and
Institut f\"ur Teilchenphysik, ETHZ, CH-8093 Z\"urich,
Switzerland}


\date{DRAFT v.~7, May 22, 2006}

\abstract{
Astrophysical sources of energetic gamma rays provide the right conditions for
maximal mixing between (pseudo)scalar (axion-like) particles and photons if
their coupling is as strong as suggested by the PVLAS claim.  This is 
independent of whether or not the axion interaction is standard at all energies or becomes suppressed in the extreme conditions of the stellar interior.  The flux of such particles through the Earth could be observed using a metre long, Tesla strength superconducting solenoid thus testing the axion interpretation of the PVLAS anomaly. The rate
of events in CAST caused by axions from the Crab pulsar is also estimated
for the PVLAS-favoured parameters. }
\maketitle

Recently, the PVLAS experiment has reported an observation of a
rotation of the plane of polarization of a laser beam passing through
magnetic field \cite{pvlas} which was claimed to be compatible with the
existence of a new (pseudo)scalar particle with a mass of $m\sim 10^{-3}$
eV and an inverse coupling to photon $M\sim 10^5$ GeV. This is unexpected
since experiments such as CAST~\cite{CAST} have seemingly ruled out this
region of parameter space. Some authors have tried to explain this
discrepancy by suggesting alternative models for the pseudoscalar
in which its effective coupling to photons is suppressed in the extreme
conditions of the stellar interior~\cite{cooked}. Alternative explanations
of the effect by means of particles carrying very small electric
charge~\cite{millicharged} are disfavoured~\cite{Ringwald:which}
by preliminary PVLAS
data and severely constrained (if not ruled out) by
the limits on the millicharged particles~\cite{millicharged:limits}. Note
that the quantitative results of the CAST experiment depend on the model
of the solar interior while the PVLAS claim is based on relatively low
statistics. Doubts have also been raised over the latter result due to
possible systematic experimental effects from the rotation of the magnet
\cite{hep-ph/0702135}. The apparent discrepancy between the two
experiments calls for more model-independent tests both in laboratory
experiments~\cite{Ringwald:which,ringwald} and in gamma-ray
astronomy~\cite{2pulsar,FRT,Raff}. Here we suggest that if new axion-like
particles exist which have a coupling to two photons as strong as that suggested by the PVLAS claim, there should be a flux of these particles through
the Earth coming from conversion of energetic gamma-rays emitted by
astrophysical sources to axions in the magnetic fields of the sources
themselves.
Since the conditions (temperature, density and average momentum transfer)
in such typical sources are much closer to those in the laboratory rather than the stellar interior,
this flux is guaranteed in any model which explains the PVLAS result with
a new (pseudo)scalar. We argue that this flux can be detected in a
laboratory experiment by using a superconducting solenoid surrounded by an
electromagnetic calorimeter.

For definiteness, let us consider the
Lagrangian density of the photon-pseudoscalar system\footnote{The
consideration and quantitative results for a scalar particle are
similar, though other effects may be important in that case.},
$$
\mathcal{L}=\frac{1}{2}(\partial^\mu a\partial_\mu a-m^2 a^2)
-\frac{1}{4}\frac{a}{M}F_{\mu\nu}\widetilde
F^{\mu\nu}-\frac{1}{4}F_{\mu\nu}F^{\mu\nu}
$$
where $F_{\mu\nu}$ is the electromagnetic stress tensor and $\widetilde
F_{\mu\nu}=\epsilon _{\mu \nu \rho \lambda }F_{\rho \lambda }$ its dual,
$a$ the pseudoscalar (axion) field, $m$ the axion mass and $M$ is its
inverse coupling to the photon field. The coupling of the photon and
pseudoscalar fields in this way means that a photon has a finite
probability of mixing with its opposite polarisation and with the
pseudoscalar in the presence of an external magnetic field
\cite{fromRaf1,sikivie,raffelt}.
The
probability of the $a\to \gamma$ oscillation after traveling the distance
$L$ in the constant magnetic field with perpendicular component $B$ is
$$
P=\frac{4 \Delta _M^2}{\left(\Delta _p-\Delta_m \right)^2+4 \Delta_M^2
} \sin^2\left(
\frac{1}{2}{L \Delta_{\rm osc}}
\right),
$$
where
$$
\begin{array}{rcccl}
\displaystyle
\Delta_{M}
&=&
\displaystyle
\frac{B}{2M}
&=&
\displaystyle
7 \!\cdot\! 10^{-7}\left(\frac{B}{4~{\rm T}}\right)
\left(\frac{10^{5}~{\rm GeV}}{M}\right){\rm cm}^{-1}\!,\\
\displaystyle
\Delta_m
&=&
\displaystyle
\frac{m^2}{2 \omega}
&=&
\displaystyle
2.5 \!\cdot\! 10^{-11}\left(\frac{m}{10^{-3}~{\rm eV}}\right)^{\! 2} \!
\left(\frac{1~{\rm GeV}}{\omega}\right){\rm cm}^{-1}\!,\\
\displaystyle
\Delta_p
&=&
\displaystyle
\frac{2\pi\alpha n_e}{ \omega m_e}
&=&
\displaystyle
3.6 \! \cdot \!
10^{-4} \left(\frac{n_e}{10^{22}~\rm{cm}^{-3}}\right)
\!\left(\frac{1~{\rm GeV}}{\omega}\right){\rm cm}^{-1}\!,
\end{array}
$$
$$
\Delta_{\rm osc}^2=\left(\Delta_p-\Delta_m \right)^2+4 \Delta_M^2,
$$
$n_e$ is the electron density,
$m_e$ is the electron mass, $\alpha $ is
the fine-structure constant and $\omega $ is the photon (axion) energy.

The discussion of Ref.~\cite{FRT}, based upon the Hillas plot presented
there, suggests that for the PVLAS-favoured parameters, the maximal mixing
between photons and axions takes place inside the source for most of the
astrophysical objects which emit gamma rays.
This fact does not depend on details of the emission mechanism and on
the models of the source; it is based entirely on the information about
the geometrical size of the objects and the magnetic fields in their outer
regions, obtained from astronomical observations.
This maximal mixing means that if the flux $F_\gamma $ of gamma rays is
detected from the source, it should be inevitably accompanied by a flux
$F_a=F_\gamma/2$ of axions of the same energy, if such strongly coupled axions exist (the
factor $1/2$ is due to complete mixing between two photon and one axion
polarizations).

Let us concentrate first on the gamma rays with energies $E\gtrsim
10$~keV and estimate the contribution of various astrophysical sources
to the axion flux. The signal at these energies would be dominated by pulsars and gamma-ray bursts since other sources, e.g.\ active galactic nuclei, contribute only at high energies ($E\gtrsim 10$~MeV) where fluxes are too low to be detected in a realistic experiment of the type we discuss.

{\bf Pulsars}. For a typical magnetosphere of a neutron star we assume
the magnetic field $B\sim10^{13}$~G at lengths $L\sim 10$~km. Due to such
extreme magnetic fields, the conditions for maximal mixing would be fulfilled at energies as low as $E\gtrsim 10^{-4}$~eV.
To estimate the fluxes of axion like particles at $E\gtrsim
100$~MeV, we use EGRET data~\cite{3EG}; there are five pulsars detected which emit such hard gamma rays (see Table~\ref{table:3EGpsr}).
\begin{table}
\begin{center}
\begin{tabular}{ccc}
\hline\noalign{\smallskip}
Name & Flux at $E>100$~MeV, & Spectral index\\
& $10^{-8}$~photons/cm$^2$/s& $\alpha $\\
\noalign{\smallskip}\hline\noalign{\smallskip}
Crab & 226 & 2.19\\
Vela & 834 & 1.69\\
Geminga & 353 & 1.66\\
1055$-$52 & 33 & 1.94\\
1706$-$44 & 112 & 1.86\\
\noalign{\smallskip}\hline
\end{tabular}
\end{center}
\caption{
\label{table:3EGpsr}
Five EGRET pulsars.
}
\end{table}
The spectral index $\alpha $ is determined as
$$
\frac{dN}{dE}\propto E^{-\alpha} ,
$$
where $dN/dE$ is the flux and
100~MeV$\le E \le$~10~GeV is the photon energy.

The fluxes of these five objects at lower energies
were extrapolated from 100~MeV down using the spectrum of the Crab pulsar (see e.g.\ Fig.~5 of Ref.~\cite{Crab} from which we find $\alpha \approx 2.35$ for
100~keV$\le E \le$~100~MeV).

The contribution of other pulsars, important at soft
gamma-ray energies only, was estimated with the help of the INTEGRAL reference
catalog~\cite{INTrefCat}. In that reference, the sources' spectra are
classified in different ways. The three relevant spectral models are
the following:

WP (``wabs powerlaw''):
$$
S(E)=w(E) A \left(\frac{E}{E_0}   \right)^\Gamma,
$$
for $\Gamma=2$.

WHP (``wabs highcut powerlaw''):
$$
S(E)=w(E) A \left(\frac{E}{E_0}   \right)^\Gamma
\left\{
\begin{array}{cl}
\exp\left(\displaystyle \frac{\displaystyle E_{\rm cut}-E}{\displaystyle
E_{\rm fold}} \right), ~~~
&E>E_{\rm cut};\\
~\\
1,
&E<E_{\rm cut},
\end{array}
\right.
$$
for $\Gamma=1$, $E_{\rm cut}=10$~keV, $E_{\rm fold}=15$~keV.

WC (``wabs cutoff''):
$$
S(E)=w(E) A \left(\frac{E}{E_0}   \right)^\Gamma
{\rm e}^{-E/E_{\rm cut}},
$$
for $\Gamma=1.7$, $E_{\rm cut}=10$~keV.

For all models, $w(E)$ is the Galactic absorption factor which is negligible at the level of approximations made in this paper and hence we set it equal to unity within our precision, $S(E)$ is the spectral energy distribution and
$A$ is given in Table~\ref{table:IGRpsr}, where these
pulsars are listed, for $E_0=1$~keV.
\begin{table}
\begin{center}
\begin{tabular}{ccc}
\hline\noalign{\smallskip}
Name & Model & $A$, photons/cm$^2$/keV\\
\noalign{\smallskip}\hline\noalign{\smallskip}
AX J0051$-$722    &  WHP & 4.17$\cdot 10^{-3}$ \\
RX J0052.1$-$7319 &  WHP & 3.85$\cdot 10^{-5}$ \\
SMC X$-$2         &  WHP & 4.30$\cdot 10^{-3}$ \\
AX J0058$-$720    &  WHP & 1.69$\cdot 10^{-4}$ \\
RX J0059.2$-$7138 &  WHP & 2.58$\cdot 10^{-2}$ \\
AX J0103$-$722    &  WHP & 6.51$\cdot 10^{-5}$ \\
AX J0105$-$722    &  WHP & 1.60$\cdot 10^{-4}$ \\
PSR B0628$-$28    &  WC  & 5.01$\cdot 10^{-2}$ \\
PSR B0656+14      &  WP  & 1.59$\cdot 10^{-3}$ \\
PSR B1509$-$58    &  WP  & 3.64$\cdot 10^{-1}$ \\
AX J1740.2$-$2848 &  WHP & 1.70$\cdot 10^{-4}$ \\
PSR J1844$-$0258  &  WHP & 1.60$\cdot 10^{-3}$ \\
PSR B1951+32      &  WP  & 2.75$\cdot 10^{-2}$ \\
\noalign{\smallskip}\hline
\end{tabular}
\end{center}
\caption{
\label{table:IGRpsr}
Pulsars from Ref.~\cite{INTrefCat} except for those listed in
Table~\ref{table:3EGpsr} and one giving negligible contribution. }
\end{table}

{\bf Gamma-ray bursts}. The magnetic field in a gamma-ray burst (GRB)
is $B\sim 10^9$~G in a region of $L\sim 10^7$~m, so the resonant mixing
happens for $E\gtrsim 1$~eV. As an approximation we suppose that all GRBs
have the same spectra which differ only by the peak fluxes.
A useful compilation of models for the GRB fluxes may be found in
Ref.~\cite{GRBhtml}.
For the spectral energy distribution of a GRB, we use the Band
function~\cite{BandFunction}:
$$
B(E)=\left\{
\begin{array}{l}
\bar{A} \left(\displaystyle E\over \displaystyle 100~{\rm keV}
\right)^\alpha \exp \left[-(2+\alpha ){\displaystyle E\over\displaystyle
E_{\rm peak}}   \right], \\
\hfill {\rm if }~ E<E_{\rm
break}\equiv{\displaystyle \alpha -\beta \over \displaystyle 2+\alpha }
E_{\rm peak}, \\
~\\
\bar{A} \left(\displaystyle E\over \displaystyle 100~{\rm keV}
\right)^\beta {\rm e}^{\beta -\alpha}
\left[{\displaystyle \alpha -\beta \over \displaystyle
2+\alpha} \, {\displaystyle E_{\rm peak} \over \displaystyle 100~{\rm
keV}}   \right]^{\alpha -\beta }, \\
\hfill {\rm if }~ E>E_{\rm break}
\end{array}
\right.
$$
We use the mean values of parameters seen by BATSE as given in
Ref.~\cite{BATSEspectra}:
$$
\alpha =-1;
$$
$$
\beta =-2.25;
$$
$$
E_{\rm break}=250~{\rm keV}.
$$
$\bar A \approx 0.01~{\rm keV}^{-1}$ normalizes the Band function to one,
$\int\limits_{50~{\rm keV}}^{300~{\rm keV}} B(E) dE=1$.

To determine the dimensionfull coefficient in front of the Band function,
we need to sum up the intensities of all GRBs for a given period of time.
We use the distribution of the peak count rates of BATSE bursts from
Ref.~\cite{Stern}.
The data used is for the energy band between 50~keV
and 300~keV, that is why we took this band in the normalisation.
We performed the integration of the histogram in
Fig.~23 of Ref.~\cite{Stern} and obtained approximately
$10^5$~photons/cm$^2$/s/year for the sum of peak count rates of all
BATSE-detected bursts. The peak count rate is 0.75 times the peak
flux~\cite{Stern}. We have to correct also for the BATSE exposure (GRBs
which happened when BATSE did not look at that part of the sky). According
to Ref.~\cite{BATSE}, the exposure correction for 50$\div$300~keV is
0.480.

Finally, we have to relate the peak count rate and the total energy of a
GRB. To this end, we use the temporal development of the spectrum. A
universal parametrisation for it reads~\cite{Fenimore,Norris}:
$$
I(t,E)=A \left\{
\begin{array}{l}
\exp\left[-\left(\frac{\displaystyle |t-t_0|}{\displaystyle \sigma _r(E)}
\right)^\nu  \right],
~~~~~~ t\le t_0;\\
\exp\left[-\left(\frac{\displaystyle |t-t_0|}{\displaystyle \sigma _d(E)}
\right)^\nu  \right],
~~~~~~ t> t_0,
\end{array}
\right.
$$
where
$$
\sigma _d(E)=0.75 \left(\ln 2 \right)^{-1/\nu } W_0
\left(\frac{E}{20~{\rm keV}} \right)^{-0.4},
$$
$$
\sigma _d(E)=0.25 \left(\ln 2 \right)^{-1/\nu } W_0
\left(\frac{E}{20~{\rm keV}} \right)^{-0.4},
$$
and we use mean values of parameters: peakedness $ \nu
=1.44$ and full width at half-maximum of the pulse $ W_0
=0.8$~s. The
dimensionality of $A$ is photons/cm$^2$/s. Note that the Band function
$B(E)$ gives the spectrum integrated over time, while at each moment, the
flux per unit time per unit energy is $I(t,E)B_1(E)$, where $B_1(E)$ is
the same Band function but with different parameters, $\alpha _1=\alpha
+0.4$, $\beta _1=\beta +0.4$. The peak flux is given by
$$
\mbox{Peak flux} = \left.
\int\limits_{50~{\rm keV}}^{300~{\rm keV}}
I(t,E)B_1(E) dE \right|_{t=t_0}=A.
$$
To relate $A$ to the coefficient in the integral spectrum, we integrate
the flux over time (the dimensionality of the following equation is
photons/cm$^2$/keV):
$$
\int_0^\infty \! dt\, I(t,E) B_1(E)=A_0 B(E),
$$
where
$$
A_0=A \left(\ln 2 \right)^{-1/\nu }W_0 \left(\frac{100~{\rm keV}}{20~{\rm
keV}}\right)^{-0.4}
\int_0^\infty {\rm e}^{-x^\nu }\, dx.
$$
We finally obtain the spectrum of our typical GRB with peak count rate
$P$ as
$$
0.66\,\frac{P}{\mbox{photons/cm}^2/{\rm
s}}B(E)~\frac{\mbox{photons}}{{\rm cm}^2 \cdot {\rm keV}}.
$$
In our assumption that all GRBs have the same spectra up to $P$, the
total flux is obtained by taking $P=10^5$~photons/cm$^2$/s/year and
dividing by the exposure factor, which results in the
contribution of all GRBs of
$$
4.4 \cdot 10^{-3} B(E)  ~\frac{\mbox{photons}}{{\rm cm}^2\cdot{\rm
s}\cdot {\rm keV}}.
$$

\begin{figure}
\begin{center}
\includegraphics[width=0.99\columnwidth,angle=0]{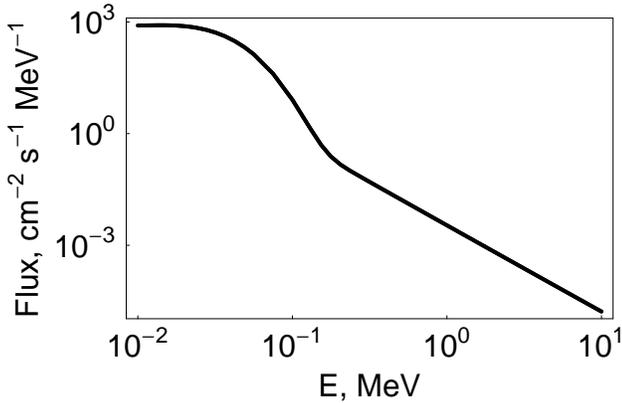}
\caption{\it
Expected axion flux from all astrophysical sources versus energy.
}
\label{fig:flux}
\end{center}
\end{figure}

The total expected flux of astrophysical axions from pulsars and GRBs is
plotted, as a function of energy, in Fig.~\ref{fig:flux}.

Let us now consider possible ways in which these axion-like particles ($a$)
could be detected in a laboratory experiment. One of the ideas of detection
proposed a long time ago \cite{bober} for searching of solar axions is the
following.
If $a$ is a long-lived particle, the flux of energetic $a$'s  would
penetrate the Earth atmosphere without significant attenuation  and would
be observed in a detector via the inverse Primakoff effect, namely in the
process of interaction of (pseudo)scalars with virtual photons from the
 static magnetic field of a superconducting magnet. An example of this
kind of experiment on high-energy axion-photon conversion may be found in
Refs.~\cite{nomaxion-gnin}.

\begin{figure}
\begin{center}
\mbox{\hspace{-.0cm}\epsfig{file=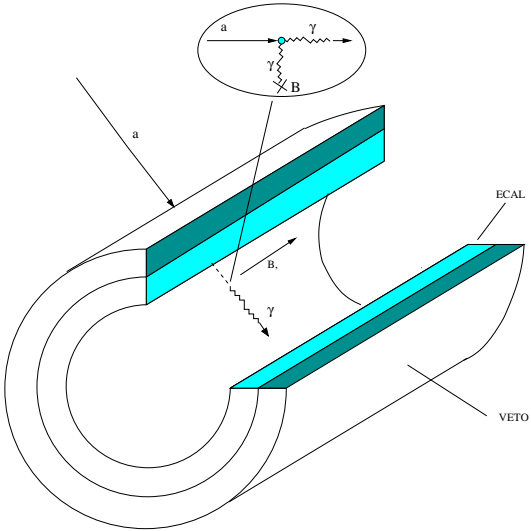,height=70mm,width=70mm}}
\end{center}
\caption{\em Schematic layout of the high energy cosmic
axion experiment. The main elements of the setup relevant
 for the axion search are illustrated.}
\label{fig:setup}
\end{figure}

The sketch of the experiment we propose is shown in Fig.~\ref{fig:setup}.
The main design feature of the detector  is
the presence of a volume $V$ of a high ($\simeq$ 1 T)
magnetic field $B$ with its internal surface covered with an electromagnetic
calorimeter (ECAL) and surrounded by a  VETO detector.
 The ECAL
detects photons with energy $E_{\gamma} \gtrsim 10$~keV.
The VETO  serves for
efficient supression  of environmental and cosmic backgrounds.
The experimental signature of $a \rightarrow \gamma$ conversion is a single
energetic  photon detected in the ECAL (either through the photo electric effect or through Compton scattering) which will not be accompanied by any energy deposition in the VETO detector.
Since the conversion happens due to the presence of the  magnetic field,
one could also search for it in the detector by comparing
event rates in the ECAL taken with and without the magnetic field.

\begin{figure}[t!]
\begin{center}
\includegraphics[width=1.02\columnwidth,angle=0]{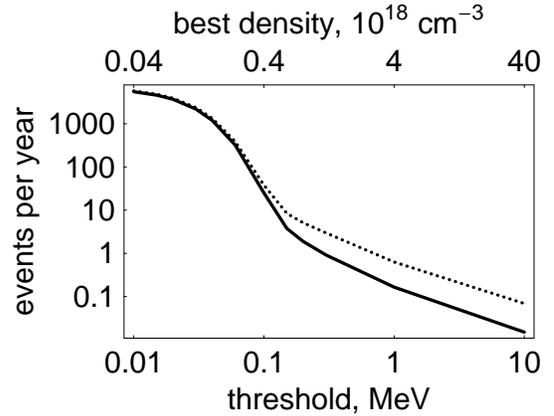}
\caption{\it
Number of events per year versus energy threshold in a
cylinder of 1~m length, 1~m radius permeated by a 1~Tesla magnetic field.
The lower curve assumes zero electron density inside the detector while the upper curve assumes an electron density chosen to bring the mixing length down to the size of the detector (shown above the plot). }
\label{fig:events}
\end{center}
\end{figure}

The statistical limit on the sensitivity of the experiment searching for
cosmic $a \to \gamma$ conversion  scales roughly as $B^2\cdot V$
\cite{bober}. Thus, to improve the sensitivity, large volume and strong
magnetic field are required. Since the gamma-ray (and potential axion)
spectra from the astrophysical objects are steep, see Fig.~\ref{fig:flux},
the detection of low-energy recoil electrons in
the ECAL is also crucial for the improvement of the sensitivity of the
search. As an example, Fig.~\ref{fig:events} presents the number of events
per year as a function of the detection energy threshold in a detector of
1m length and 1m radius filled with 1~Tesla magnetic field. The signal in such a detector will be optimised not for maximal mixing, as the mixing length in that situation will be much larger than the size of the detector.  The presence of a non-zero electron density allow one to tune the mixing length so that the probability is increased, despite the fact that the mixing angle will be reduced.  It is seen
that at lower energies one expects thousands of events.  Note that the
majority of axions at these energies arrives from pulsars; as is seen
from Tables~\ref{table:3EGpsr},~\ref{table:IGRpsr}, there are only a few
strong sources among them, hence the background might be reduced if the
gamma direction can be determined. In any case, such matters would have to be considered in detail if one had to choose the actual orientation of the detector.

A possible approach which could satisfy the requirements discussed above
is to use a  massive liquid argon time
projection chamber (TPC) as the ECAL. The external region of the chamber
could also be used as the VETO. An example of such a detector is the one
being developed in the framework of the ArDM project for the search of the
Dark Matter \cite{ardm}. This experiment is designed to run with an effective target mass of almost 1 ton and is capable of measuring energy depositions as low as 10 keV.

The first results from a liquid argon TPC in a magnetic
field look quite promising \cite{lar}.
A small liquid argon Time Projection Chamber (LAr TPC) was operated for
the first time in a magnetic field of 0.55 Tesla. The imaging properties
of the detector were not affected by the magnetic field. In a test run
with cosmic rays, a sample of through going and stopping muons was
collected. The chamber with the readout electronics and the experimental
setup are described in Ref.~\cite{lar}, where examples of reconstructed
and analyzed events are presented.

 The significance of the
 $a$-particles discovery  with the proposed   detector scales as
\cite{bk}:
$$
S= 2(\sqrt{n_s + n_b}-\sqrt{n_b}) \simeq n_s/\sqrt{n_b},
$$
where  $n_s$ and
$n_b $ are the number of detected signal and  background events, respectively.
Thus, assuming the
PVLAS value for the axion-photon coupling, requiring that $S\gtrsim 3$
and assuming $n_s \simeq 10^2$ at the ECAL energy threshold $\simeq 0.1$ MeV (see Figure \ref{fig:events}) a  background level of  $n_b\lesssim 3$ event/day
has to be achieved, which seems to be quite realistic \cite{ardm}.

One of the main sources of the background to $a\rightarrow \gamma$
events in the proposed experiment  is expected from the neutrino processes
with a significant electromagnetic component in
the final state and with no significant energy deposition in the VETO.\
The expected amount of neutrino  background events
 can be  evaluated using Monte Carlo simulations.\
Assuming a total mass of the active part of the detector to be $\simeq 1$
ton, the neutrino background is estimated to be  $n_b\lesssim 1$ event/day. Note
that the above considerations give the correct order of magnitude for the
sensitivity of the proposed experiment and may be strengthened and
extended to the low recoil energy $E \gtrsim$ 10 keV by more accurate and
detailed Monte Carlo simulations.

Let us turn now to the lower energies. As it has been pointed out above,
pulsars provide necessary conditions for maximal axion-photon mixing at
$E\gtrsim 1$~eV. We note that, assuming the PVLAS-favoured couplings,
keV-energy axions should be detected by CAST if it is pointed to the Crab
pulsar. Such pointing was indeed performed~\cite{CASTreport} but the study
has not yet been published. Given the Crab pulsar flux of about 2 photons/cm$^2$/s
in the energy interval 1~keV$\lesssim E\lesssim 14$~keV~\cite{Crab}, one
expects (for the PVLAS-favoured parameters) $\sim 0.05$ events per 24 hours
of observation by CAST (to be compared with $\sim 7$ events expected --
and not observed -- from the Sun for the {\em minimal} axion model with
$M\sim 10^{10}$~GeV). This flux can be tested by the experiment on the
time scale of a year, given 3 hours of pointing per day. Other pulsars have
much lower fluxes and give negligible contributions.

To summarize, the presence of an axion-like particle with PVLAS-favoured
parameters, no matter what kind of interactions it has in the extreme
conditions (e.g.\ inside the Sun), can be tested with dedicated
experiments using particle-physics detectors of meter-scale size and
Tesla-scale magnetic fields.
At $E\gtrsim 100$~keV, several dosen
events per year are expected in a $\sim 1$~m$^3$, $\sim 1$~T detector. At
the energies of a few keV, CAST (properly pointed) may detect about one
event from the Crab pulsar per $\sim 480$ hours observational time.
Non-observation of the predicted number of events would disfavour the
axion explanation of the PVLAS anomaly.

{\bf Acknowledgments}
We are grateful for discussions with V.~Kuzmin, S.~Popov, V.~Rubakov,
S.~Sibiryakov, B.~Stern, I.~Tkachev and K.~Zioutas.
This work was supported in part by
the Swedish Research Council (Vetenskapsr\aa det) and the Perimeter
Institute (MF), by the Marie Curie Incoming International Fellowship of
the EC, RFBR grant 04-02-16386 and the RAS Program ``Solar activity''
(TR), by the Russian Science Support Foundation fellowship, grants INTAS
03-51-5112, RFBR 07-02-00820 and NS-7293.2006.2 (ST), by the RFBR grant
07-02-00256 (NK). SG and ST thank ETH and CERN, respectively,  for
hospitality and support at the initial stages of the work.

\end{document}